\documentclass[]{aa}
\usepackage{graphicx}
\begin{document}

\title{First detection of dust clouds around R~CrB variable stars
\thanks{Based on observations collected
with the VLT/UT4 Yepun telescope (Paranal Observatory, ESO, Chile)
using the NACO instrument (program ID 71.D-0543A)}}


\author{P. de Laverny
       \and
        D. M\'ekarnia 
          }


   \institute{Observatoire de la C\^ote d'Azur, Dpt Cassiop\'ee, 
             CNRS-UMR 6202, 
             BP 4229, 06304 Nice Cedex 4, France \\
              \email{\{laverny;mekarnia\}@obs-nice.fr}
             }

   \date{Received 6 September 2004 /Accepted  23 October 2004}

   \titlerunning{Dust clouds around R~CrB variables stars}


   \abstract{
From VLT/NACO diffraction-limited images of RY~Sgr,
we report the first direct detection of heterogeneities
in the circumstellar envelope of a R~Coronae Borealis variable star.
Several bright and very large dust clouds are seen  in various directions
at several hundred stellar radii from RY~Sgr, revealing 
high activity for the ejection of stellar material
by R~CrB variables.
These observations do support the current
   interpretation that optically thick dust clouds are formed 
around the surface of this type of variable stars
and, when passing between the star and the observer, produce
the huge and sudden declines characterizing these objects in visible light.
This is the
first direct confirmation of a scenario proposed about 70~years ago.
   }

   \maketitle

\section{Introduction}

{\it ``Le R Coridini (protebbero), per qualche motivo,
cingersi (....) di gas oscuranti emessi a tratti
da queste stelle stesse''}. In this way, Loreta (\cite{loreta})
was the first to propose that R~Coronae Borealis type variables
(R~CrB, hereafter) might eject absorbing material responsible of
the huge brightness declines (or fadings) characterizing these stars
in visible light. Then, O'Keefe (\cite{okeefe}) showed that
this ejected material could condense at small distances from the
central star and form obscuring clouds
rich in carboneous compounds. These clouds would then gradually
dissipate as they are driven away by radiation pressure.

For more than 70 years, this scenario for the interpretation
of the light variations of R~CrB variable stars has been widely accepted:
the fadings are believed to be caused by
obscurations of the stellar surface by newly formed dusty and
optically thick clouds.
These stars indeed exhibit drastic and erratic variabilities.
Their visual lightcurve is  characterized by unpredicted
decreases of up to 8 magnitudes, in a time-scale of weeks.
The return to normal light is much slower and can last up to several months
(see  Clayton \cite{clayton96}). 
There are several observational evidences 
in favor of this scenario (see Clayton \cite{clayton96}, for
more details). Among them, the fact that (i) the dust 
consists primarily as amorphous carbon particles, (ii) mass-loss rates are 
as large as 10$^{-6}$M$_{\odot}$/year and episodic, with timescales
of a few months, (iii) mass-loss is driven by fast winds (Clayton et al. 
\cite{clayton03}),
(iv) the dust may form only over a small solid
angle of the stellar surface (perhaps the cool regions above 
large convective cells, as first proposed by Wdowiak \cite{wdowiak}) 
or is ejected in some specific directions 
(Clayton et al. \cite{clayton97}), (v) polarimetric observations may reveal
the presence of permanent clumpy non-spherical dust shells (Clayton et al. 
\cite{clayton97} and Yudin et al. \cite{yudin}),
(vi) larger polarizations are seen
during declines revealing more scattering particles 
between the star and the observer (see references in 
Clayton \cite{clayton96}), 
(vii) characteristic time-scales of the
light variations are compatible with the formation of dust clouds
close to the stellar photosphere and their dilution in the outer regions
(Hartmann \& Apruzese \cite{hartmann} and Zubko \cite{zubko}).

However, in spite of the above evidences, 
the dusty environment close
to the star remains almost unknown: no direct detection
of the suspected heterogeneities
in the dust distribution in the circumstellar envelope of a R~CrB variable
have up to now been achieved. 
Observations of these inner layers in R~CrB were reported by Ohnaka et al.
(\cite{ohnaka03} and previous references) but no significant 
deviation from circular geometry was detected (perhaps because of
the rather small dynamical range of their observations,
inherent to the technique adopted). In this letter, we
present the first direct detection of the presence of 
dusty clouds around RY~Sgr, 
the brightest R~CrB variable in the southern hemisphere.

\section{Observations}

Observations of RY~Sgr were performed in service mode in May
and September 2003 using the adaptive optics system NACO at the
Nasmyth-B focus of the ESO/VLT fourth 8-m telescope unit Yepun located at Cerro
Paranal, Chile. NACO, which is an association of the adaptive optics
system NAOS (Rousset et al. \cite{rousset}) and the spectro-imager CONICA
(Lenzen et al. \cite{lenzen}), provides
diffraction-limited images in the
near-infrared (1-5$\mu$m) spectral range 
(see http://www.eso.org/instruments/naco/).

RY~Sgr was observed using three narrow-band filters: NB~1.04
(centered at $\lambda_c$=1.04 $\pm$ 0.015$\mu$m),  
NB~2.17 ($\lambda_c$=2.166 $\pm$ 0.023$\mu$m) and
NB~4.05 ($\lambda_c$=4.051 $\pm$ 0.02$\mu$m). The pixel scale on
CONICA was respectively 13.25~mas in the 
NB~1.04 and NB~2.17 filters and 27.03~mas in
the NB~4.05 filter, adapted to the observing wavelengths. The seeing conditions were variable,
ranging from $\sim$~0.5\arcsec \ during observations at 2.17$\mu$m to
$\sim$~0.9\arcsec \ during observations at 4.05$\mu$m.
The Auto-Jitter mode was
used, i.e. that at each exposure, the telescope was moved according to a
random pattern in a 6\arcsec~$\times$~6\arcsec \ box. The sky is then
estimated from all the observations. To allow further resolution 
improvement through deconvolution, a
PSF reference star (HD~178199) was observed
immediately after each RY~Sgr observation using the same configuration 
of the adaptive optics system.  The air-masses
of RY~Sgr and of the
PSF reference star were 1.0 and 1.1 respectively
and NAOS was servoed on RY~Sgr itself. The FWHM of the PSF, estimated 
from the reference star in the different narrow-band filters, was  
0.058\arcsec, 0.068\arcsec \
 and  0.116\arcsec \ at 1.04$\mu$m, 2.17$\mu$m 
and 4.05$\mu$m respectively.   Calibration
files (flat fields and dark exposures) were acquired, following the ESO/VLT
standard calibration plan. The total
on-source integration time in the three filters were 
 80~mn, 33~mn and 3~mn respectively. The
dynamic range of the final images goes from  
 3\,300 at 2.17$\mu$m, 12\,500 at 1.04$\mu$m to 54\,000 
at 4.05$\mu$m. Table~1 summarizes the 
observational conditions. 

Standard reduction procedures were applied using self-developed 
routines. The raw images were sky subtracted, then
divided by the flat-field and corrected from hot pixels. In each
filter, the images were cross-correlated and 
aligned by sub-pixel shifting, and then combined to produce the 
final images, eliminating
discrepant points like cosmic rays. Finally, 
the images  were deconvolved with the
PSF reference star. We used the Richardson-Lucy algorithm
(Richardson \cite{richardson} \& Lucy \cite{lucy}), which appears to be
more suitable for data with a high dynamic range, such as ours. 
 Constancy of prominent features present in deconvolved images
showed that the PSF selection and the number of iterations (10-20 iterations, typically) for
the deconvolution was performed carefully and conservatively.

   \begin{table}
      \caption[]{Observations log.}
         \label{tab_log}
         \begin{tabular}[]{l c c  c }
            \hline
            \hline
            \noalign{\smallskip}
            Date &  $\lambda_c$   & Exp. time  & Seeing \\
            (UT) & ($\mu$m)       & (on-source in min) & (\arcsec) \\
          \noalign{\smallskip}
            \hline
            \noalign{\smallskip}
  2003 May 17       & 1.04 & 80 & $\sim$ 0.8  \\  
  2003 May 24       & 2.17 & 33 & $\sim$ 0.5  \\
  2003 September 17 & 4.05 & 3  & $\sim$ 0.9  \\
            \noalign{\smallskip}
            \hline
         \end{tabular}
   \end{table}

\section{Images of the circumstellar envelope of RY~Sgr}

   \begin{figure}
    \includegraphics[width=9.cm]{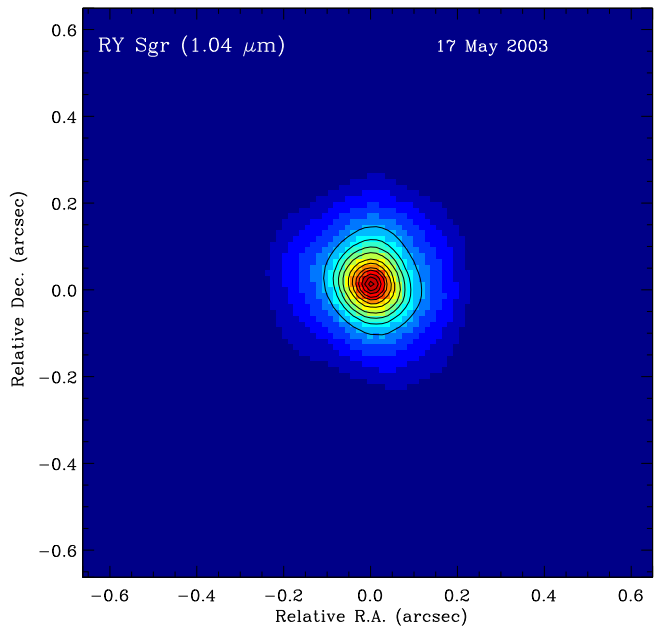}
   \caption{NACO image of RY~Sgr at 1.04$\mu\mathrm{m}$. 
   Contour levels are superimposed on the
   brightness-color log-scale of the image.
   They are 90, 70, 50, 30 , 20,
   10, 5,  2, 1, 0.5 and 0.2\% of the peak surface brightness. North
   is up, and east is to the left. The spatial resolution
   estimated from the PSF reference star is 58~mas.}
    \end{figure}

    \begin{figure*}
    \includegraphics[width=18.cm]{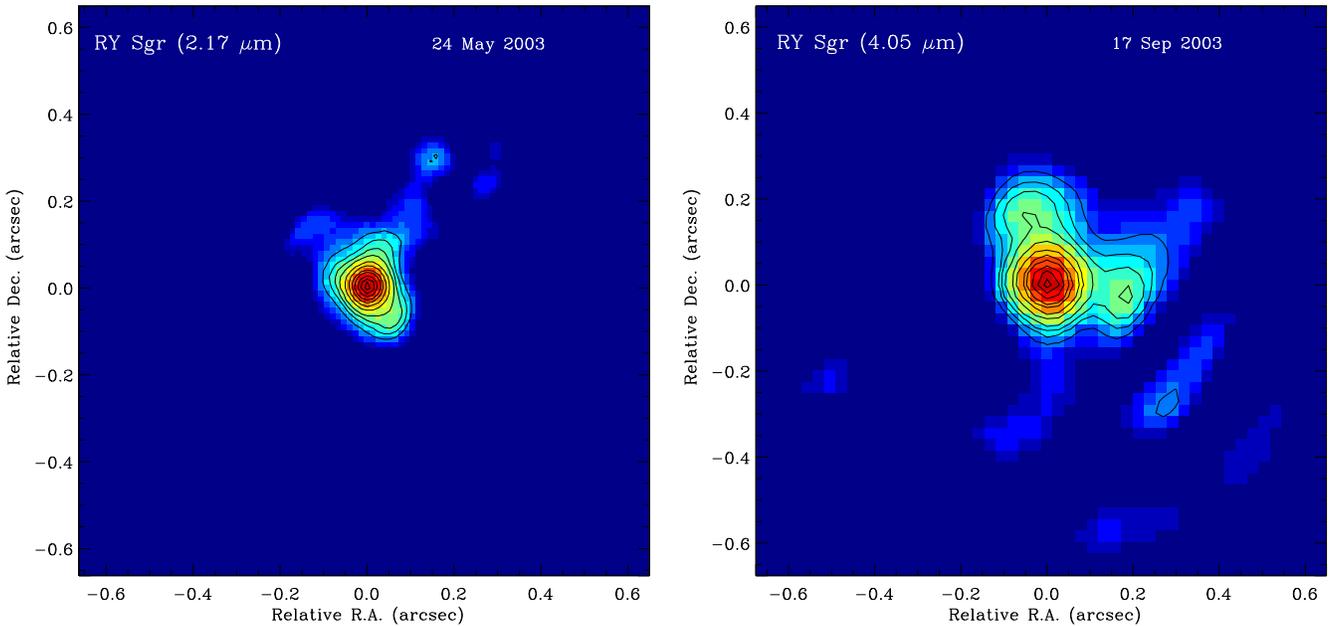}
   \caption{NACO images of RY~Sgr at 2.17$\mu\mathrm{m}$ (left) and 
   4.05$\mu\mathrm{m}$ (right, collected 4 months later). 
   Same contour levels and orientation as Fig.~1.
   The pixel scale at 4.05$\mu\mathrm{m}$ 
   is twice larger due to NACO characteristics. The spatial resolution
   estimated from the PSF reference star is 68~mas and 116~mas at 
   2.17$\mu\mathrm{m}$ and 4.05$\mu\mathrm{m}$, respectively. Note that 
   the structures seen in both images result from 
   different ejection events (see text).}
              \label{}%
    \end{figure*}

The diffraction-limited images of RY~Sgr 
at 1.04$\mu\mathrm{m}$, 2.17$\mu\mathrm{m}$  and 
   4.05$\mu\mathrm{m}$ are presented in Fig.~1 \& 2. The images are
   displayed with a log-scale for the brightness, so that details of
   the morphology at
   all flux levels can be seen. In the 1.04$\mu\mathrm{m}$ image (Fig.~1)
collected at the same epoch as the 
2.17$\mu\mathrm{m}$ one (Fig.~2, left), 
RY~Sgr is not resolved and no structures
are seen in its envelope. 
On the other hand, RY~Sgr reveals strong departures from a point-like
source at longer wavelengths since several structures are seen
in the K (2.17$\mu\mathrm{m}$) and L(4.05$\mu\mathrm{m}$) images (Fig.~2). 
In this section, we describe only the brightest detected 
heterogeneities,
i.e. those appearing brighter than 0.2\% of the peak level
and thus well above the noise level defined by
the dynamics of the images.

In the K-band, two elongations are clearly seen
towards the NW and SW directions (at PA $\sim$330$^o$  and
PA $\sim$195$^o$ respectively). These bright structures
are located at about 0.1\arcsec \ from the central star
and they appear to have a typical size similar to the one of the
central star itself.

At longer wavelengths (4.05$\mu\mathrm{m}$), the departure
from a point-like symmetry is even more evident.
The most prominent structures are two clouds as bright as 2\%
of the stellar peak. They are found at about 0.2\arcsec \ from the 
center of the image (i.e. twice further than in the K-image, revealing that
different clouds are seen in the L-band)
at PA $\sim 20 ^o$ and PA $\sim 260^o$. Their size is of the order of
0.2\arcsec. 

Finally, fainter structures are seen in both images of Fig.~2. 
These structures  could be
artifacts in K-band but they are about 10 times brighter
than the noise level in the L-band image. Therefore, RY~Sgr might be surrounded
at a given epoch by several clouds (about ten or so) located 
at distances up to 0.5\arcsec \ from its center.
Nevertheless, this more complex configuration should be confirmed 
by new independent observations.

\section{Discussion}

These images  do reveal  undoubtedly that large clouds are
present in the vicinity of R~CrB variables. 
They are seen in different directions and located at 
different distances from the central
star. 
It has to be pointed out that the clouds seen in the K and L-images
result from different ejection events. Indeed, 
an extremely large velocity (more than one order of magnitude above the
typical escape velocity observed around R~CrB variables) would be required 
to explain such a cloud motion (0.1\arcsec \
in 4 months), at the estimated distance of RY~Sgr ($\sim$ 2~kpc, see below).
We also note that the clouds appearing at $\sim$0.2" from the center
of the field in
the L-band image are not seen in K-band
because,  despite the better spatial
resolution, the dynamic range in K may not be high enough
to detect these rather cold and therefore too faint clouds.
On another hand, the structures seen in May~2003 in K-band 
are not detected in September in L-band. A possible
explanation is that the angular resolution of the L image (0.116\arcsec) is not
high enough to resolve these features that are present at about 0.1\arcsec \
from the center of the field. 
In any cases, it is very likely that these clouds 
are composed of 
dust particles since they are not detected at 1.04$\mu\mathrm{m}$
while they clearly appear at the same
epoch in K-band  where their emission is high enough to be detected.

Therefore, these observations do confirm, 
for the first time, the scenario proposed several decades
 ago by Loreta (\cite{loreta}) and O'Keefe (\cite{okeefe})
and now widely accepted. When a very optically thick dust 
cloud is ejected towards
the observer, a huge brightness decline, 
characteristic of R~CrB variable stars, is observed in visible
light. On the contrary, almost no variations are seen at longer wavelengths
where this cloud is optically thin.

The rather large number of clouds detected around RY~Sgr also reveals a high
activity for the R~CrB variable stars regarding the ejection 
of stellar material. Such large departures from spherical symmetry 
around RY~Sgr could explain why brightness declines
of R~CrB variables are not so rare (a few every 10~years, typically): 
a larger number
of ejected clouds (if the ejection is isotropic)
leads to a larger probability that one of them lies on
the line of sight. R~CrB stars with the most frequent brightness
declines would therefore be the stars ejecting material at
the larger rate. For instance, RY~Sgr exhibited about 10 declines over the 
last 50~years (from AAVSO lightcurve). That is a frequency about half 
that of the star R~CrB, which should be surrounded by a higher
number of dust clouds
and might therefore eject material at a higher rate.

In addition, not only the number of ejected clouds is large but these clouds
might be very dense and optically thick close to the stellar surface.
It has indeed to be noted that 
the mean density of the circumstellar layers,
where they are detected in the NACO images,
is a factor of
the order of $10^{6}$ lower (assuming
a density law varying as $r^{-2}$ in the envelope)
with respect to the regions very close to the star
where they are formed. 
If we assume that the density in the clouds has decreased by the same amount,
it results that an impressive quantity of material
is suddenly ejected from the stellar surface and eventually form
the dust clouds surrounding these stars.

Furthermore, the dusty clouds are detected rather far from RY~Sgr itself,
shedding new light on their dilution into the interstellar
medium.
RY~Sgr is a rather hot R~CrB variables ($T_{\rm eff}$ = 7\,000-7\,500~K,
following Asplund et al. \cite{asplund}). It could thus belong 
to the class of the brightest R~CrB with $M_V$ = -5 as revealed
by SMC \& LMC R~CrB variables (Alcock et al. \cite{alcock} and
Tisserand et al. \cite{tisserand}, assuming
that galactic R~CrB stars have similar properties). Adopting a bolometric
correction BC = -0.15 for a G0 supergiant, the radius of RY~Sgr is
approximately 60~$R_\odot$. From the adopted absolute magnitude
and the AAVSO photometry ($m_V \sim 6.4$ at maximum light), 
we estimate that RY~Sgr lies at about 1.9~kpc, yielding 
an angular radius 0.15~mas for this star. 
The dust clouds shown in Fig.~2, placed at about 0.1-0.2\arcsec ,
are therefore located between $\sim 700$-$1\,400~R_*$
and their typical radius is also close to a few hundred stellar radii.
That offers new constraints on the dilution of these clouds
in the outer circumstellar envelopes of R~CrB variables, since they still
have an important size far from the region where they were formed.
The NACO images indeed reveal that the light recovery 
of R~CrB stars in the optical would not be caused by the evaporation of the 
clouds close to the stellar surface since they are still seen
far from the central object. These clouds are indeed rather steady 
because they have been ejected a few years ago (about 5-10~years,
assuming a typical escape velocity of at least 200~km.s$^{-1}$, 
Clayton et al. \cite{clayton03}). We also note that, if the most distant
structures seen at about 0.5\arcsec \ (or more than 3\,000~$R_*$)
are confirmed, they might have been ejected a few decades ago.
This reveals that dust clouds around R~CrB variables 
move away from the star leading, if present on the line of sight,
to less obscuration of the surface. Therefore, the return to maximum light
might not be caused by the evaporation of the clouds very close to the
star.

Finally, regarding the location of the formation of the
dust clouds, we cannot
discriminate from the collected data with insufficient spatial resolution
between the  two commonly adopted
scenarios: either the dust is formed very close to the
stellar surface ($\sim 2~$R$_{\star}$) or it is formed
in regions more distant than $\sim 20~$R$_{\star}$
(see Clayton \cite{clayton96}). 

\section{Summary and conclusion}
We have presented the first detection of heterogeneities
in the circumstellar envelope of a R~Coronae Borealis variable star
owing to high spatial resolution images collected with NACO.

Several large dust clouds are found in different directions 
at several hundred stellar radii from RY~Sgr. These observations 
do support the current interpretation that the huge and sudden declines
which characterize these objects in optical light are caused by the formation of dust
clouds along the line of sight. However, the present data are not sufficiently
spatially resolved to
verify whether the dust is formed close to the stellar surface or in
more distant regions.  Future observations with higher
spatial resolutions (ideally a factor 10 or more),
also with very high dynamics, should help to better 
understand the circumstellar environment
and the evolution of these variable stars.

\begin{acknowledgements}
We thank A.P.P. Recio-Blanco for her careful reading of the manuscript
and her Anglo-Italian languages
expertise. G. Niccolini is acknowledged for comments and 
discussions on this work. The referee (G.C. Clayton) and M. Asplund 
are thanked for their valuable suggestions.  
We are also greatful to the variable star observations from the AAVSO 
International Database, contributed by observers worldwide and used 
in this research. 
\end{acknowledgements}

\end{document}